# Smart Cities and Villages: Concept Review and Implementation Perspectives in Developing Cities

Kamiba I. Kabuya, Olasupo O. Ajayi, Anotine B. Bagula

Department of Computer Science, University of the Western Cape, Bellville, 7535, Cape Town, South Africa.

**Abstract**

The "Smart City" (SC) concept has been around for decades with deployment scenarios revealed in major cities of developed countries. However, while SC has enhanced the living conditions of city dwellers in the developed world, the concept is still either missing or poorly deployed in the developing world. This paper presents a review of the SC concept from the perspective of its application to cities in developing nations, the opportunities it avails, and challenges related to its applicability to these cities. Building upon a systematic review of literature, this paper shows that there are neither canonical definitions, models or frameworks of references for the SC concept. This paper also aims to bridge the gap between the "smart city" and "smart village" concepts, with the expectation of providing a holistic approach to solving common issues in cities around the world. Drawing inspiration from other authors, we propose a conceptual model for a SC initiative in Africa and demonstrate the need to prioritize research and capacity development. We also discuss the potential opportunities for such SC implementations in sub-Saharan Africa. As a case study, we consider the city of Lubumbashi in the Democratic Republic of Congo and discuss ways of making it a smart city by building around successful smart city initiatives. It is our belief that for Lubumbashi, as with any other city in Sub-Saharan Africa, the first step to developing a smart city is to build knowledge and create an intellectual capital.

1. **INTRODUCTION**

In 2015 the United Nations (UN) set up 17 Sustainable Development Goals (SDGs) aimed at transforming our world by 2030. It was a 15 year plan geared toward ending poverty and inequality, addressing the growing climatic and environmental concerns, as well as ensuring peace and justice for all [UN, a]. Goal #11, sustainable cities and communities, aims at ensuring adequate, safe, and sustainable communities, by increasing the number of ample urban houses, efficient transportation systems and green spaces across cities. In UN [2023], it is opined that achieving this 11th SDG can translate to reducing the adverse per capita environmental impact of cities, and foster economic growth. To achieve this goal of sustainable communities, especially in Africa, Asia, and remote islands, it was suggested that steps must be taken to develop industries and infrastructure (SDG #9). This might include developing new infrastructure or upgrading and/or retrofitting older ones to make them sustainable and resource-efficient. Knowledge discovery, through scientific research and technological adoption (indigenous and otherwise), was also considered critical to the actualization of SDG #11.

Though the concept of Smart City (SC) is often associated with urban sustainable cities and revolves around technological innovations and digital transformations within such cities, this represents an oversimplification of the concept as it discounts the 'dynamics' or interactions between the components, which is key for all smart cities. Rather, the SC concept should be perceived from a more holistic and multidisciplinary perspective. The city must be seen as a connected entity of integrated systems and dynamic sub-systems. Therefore, the "smartness" of a city relies on the smartness of its subsystems, the interaction between them, and the data produced and consumed therein. This also includes the interconnecting links that emerge and the effect the city's evolution has on them. In essence, the subsystems that make up an SC can be likened to processes and variables that evolve over time; hence are dynamic in nature. This is in agreement with the concepts of "Urban Dynamics" proposed in [Wolman, 1969]. Beyond urban dynamics, a smart city can also be associated with many other concepts such as sustainability, eco-city, green space, clean energy, etc.

Though several reports claim that approximately half of the world's population lives in urban areas with this number expected to grow to 70% by 2050, this might not be the case in global southern countries. In many developing countries and small remote islands, close to 40 % of the citizenry live below the poverty line and dwell in villages and rural communities. This represents
a significant imbalance and inequality, and a major setback to the UN's SDG #10 aim of reducing inequality by 2030. With the target of 2030 already on the horizon, Smart Villages (SV) might be a potential solution to help the UN achieve its mandate of bridging the inequality gap between the wealthy urban dwellers, and those living in informal settlements and under-served remote villages globally.

This work explores the concepts of smart cities and villages, their building blocks, and various related initiatives globally. The specific contributions of this paper are:

• to find a suitable definition of a smart city and its potential correlations to other concepts. The expectation is to bridge the gap between different smart city concepts and reach a unified definition.
• drawing from the survey by Neirotti et al. [2014], we note that there is a significant gap between the number of scientific publications on SCs in Africa compared to those of other continents. With limited data and an understanding that SCs require modeling initiatives, this paper seeks to explore the feasibility of an African SC. In doing this, we propose an African
approach to SC initiatives and present plausible challenges of such implementations.
• to discuss potential opportunities for implementing SC initiatives in SubSaharan African cities using Lubumbashi, a city in the Democratic Republic of Congo (DRC), as a case study.

This paper is subdivided into seven sections, with the first being the introductory section. In the second section we elaborate on Smart Cities and related concepts. In the third section, we present the concept of Smart Village as a precursor for SC in developed nations; while in the fourth, we identify some SC initiatives around the world. In the fifth section, we compare SC initiatives in Europe and Africa, then present the Lubumbashi case study in section six and discuss implementation considerations. Finally, we conclude the paper in the seventh section and present avenues for future considerations..

## 2. THE SMART CITY CONCEPT

Of the numerous publications referenced in this paper, none claims to provide the most accurate definition of the Smart City concept. It can therefore be inferred that there is yet to be a generic/common definition of an SC. In this section, some key concepts relating to smart cities are discussed with the expectation of shedding light and aiding a better understanding of smart cities. Some of these concepts were proposed in Pichler [2017], while others are extracted from [Teipelke, 2018, Finger, 2018]. They include but are not limited to urban dynamics, ICT, sustainability, and Eco-cities. These concepts, discussed in this section, will provide a better understanding of the urban and empirical effects of the SC concept.

### 2.1 Views on Smart City

The significant evolution of SCs in recent years has been aided by the UN's push for SDG #11 and the consequent increase in a number of scientific researches, which have resulted in a better understanding of the concept. A direct consequence of this is a rise in the number of implementations and deployed projects globally. Despite these, there is still no universally accepted definition of the term 'Smart City'. This claim is buttressed by the dispersed definitions given by numerous authors of scientific literature (themed and otherwise), such as Neirotti et al. [2014], Nam and Pardo [2011]. In light of this, a definition for the smart city concept might be derived from the characteristics of SC found in different models or frameworks that exist in literature.

However, while relying on literature for a definition, the context must be taken into consideration. From a contextual view, SC and indeed Smart Village (SV) are viewed differently across nations and citizens. This contextualization, though paramount, might become a limiting factor if a global view or generic definition of the concept is required. One such effort, which sought to capture both a contextual and global view of SC is the work done by Varghese [2016b]. Though the author gave a contextual definition of SC from an Indian perspective, they did not lose sight of the generic and global vision of SC as a component of the 21st century. This generic vision usually includes enhanced economy, ICT innovation, improved quality of life, mobility and transport, education, and health care.

From a historical standpoint, SC can also be considered as an evolutionary concept. As early as the 1980s, there were talks of the "Information City". This was defined as a digital environment where information is collected from local communities for dissemination to the public via web portals [8]. In the 1990s, the evolution of ICTs and the Internet brought up the terms "Digital

City" and "Virtual City". These terms are often used as synonyms and include communities of urban spaces connected by ICTs for data exchange and sharing. Over the years, the concept of the Digital City evolved towards a "Ubiquitous City" or U-City – a term adopted in 2007 by South Korea [Tadili and Fasly, 2019]. U-city would later become "Intelligent City" and then "Smart City" as seen today. This first evolution of the SC concept had a one-dimensional connotation centered around technological development.

The current vision of an SC is multi-dimensional. The vision consists of embedding into the "smart" concept, all factors and components of sustainability and intelligence into a city irrespective of its size, location, and those who live in it. These embedded components, in our view, provide a clear distinction and addresses the confusing issue of the difference between a Digital and a Smart City. It follows from this definition that the "Smart City", ipso facto, integrates the "Digital City" concept, though the two concepts are different contextually. In the past two decades of terminological evolution, a "Digital City" is often attributed to the progress seen in terms of ICT [Nam and Pardo, 2011]. On the other hand, the term "Smart" implies some form of "intelligence". It encompasses scientific research related to innovations in all urban dimensions of a city, including education, health, safety, identification of the population, the economy and its transactions, well-being and habitat, transport and mobility, etc. This reveals that "Smart City" is a complex concept that involves many dynamic dimensions at the same time. These dimensions might include: i.) human or intellectual capacity – the density of knowledge and resources skills available within a city; ii.) technological – ICT innovations including networks, Big Data, Analytics, Internet of Things (IoT), Cloud Computing, etc.; iii.) institutional – all the organizational capital of communities, organizations, and private and public institutions under the leadership and superintendence of humans using ICT [Nam and Pardo, 2011, Dameri, 2017]. Expressions such as "Smart Technology", "Smart people" and "Smart Community" emanated from these dimensions.

Researchers have proposed various frameworks as guides to defining SC. For instance, in 2012, Cohen [2012] presented the characteristics of the SC in the form of a six-dimensional framework, consisting of a Smart Economy, Smart People, Smart Mobility, Smart Living, Smart Environment, and Smart Governance. The framework has become a benchmark in most SC projects around the world. He draws from the SC Wheel Framework shown in Fig. 1a and incorporates economy, mobility, and environment into the initial three dimensions given in [Nam and Pardo, 2011, Dameri, 2017]. Integration and interdependence among these dimensions is a requirement for smartness, hence incorporated in them; as no dimension can be said to be smart if it works autonomously. Each dimension of the "smart city" is therefore correlated to the other, each projecting into the plane of the other where it has a role (or more) to play.

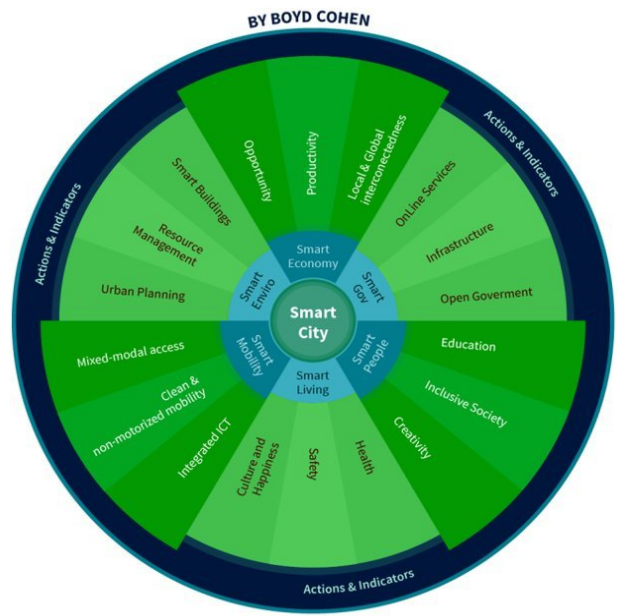

(a): Cohen's Smart City wheel [Cohen, 2012]

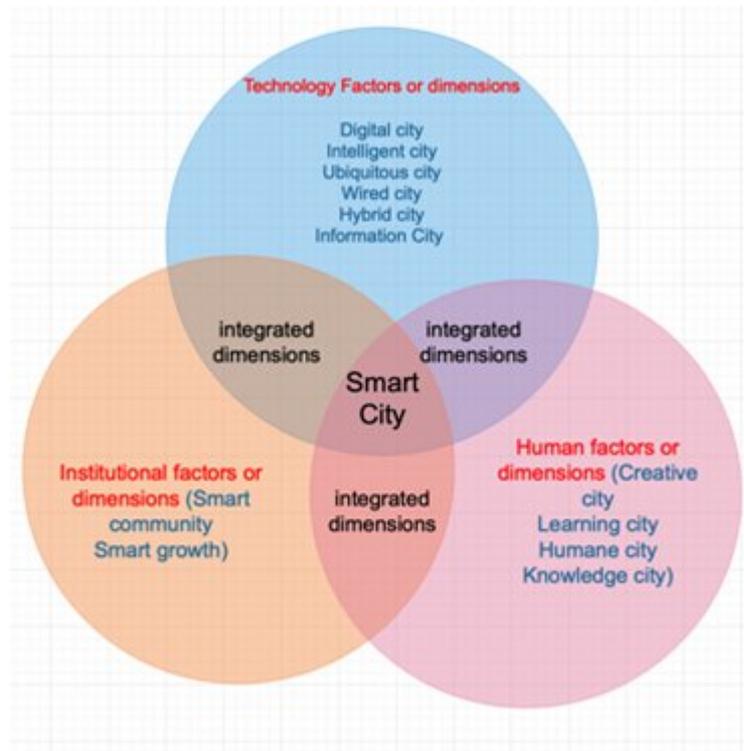

(b) Pardo's Smart City Conceptual Model [Nam and Pardo, 2011]

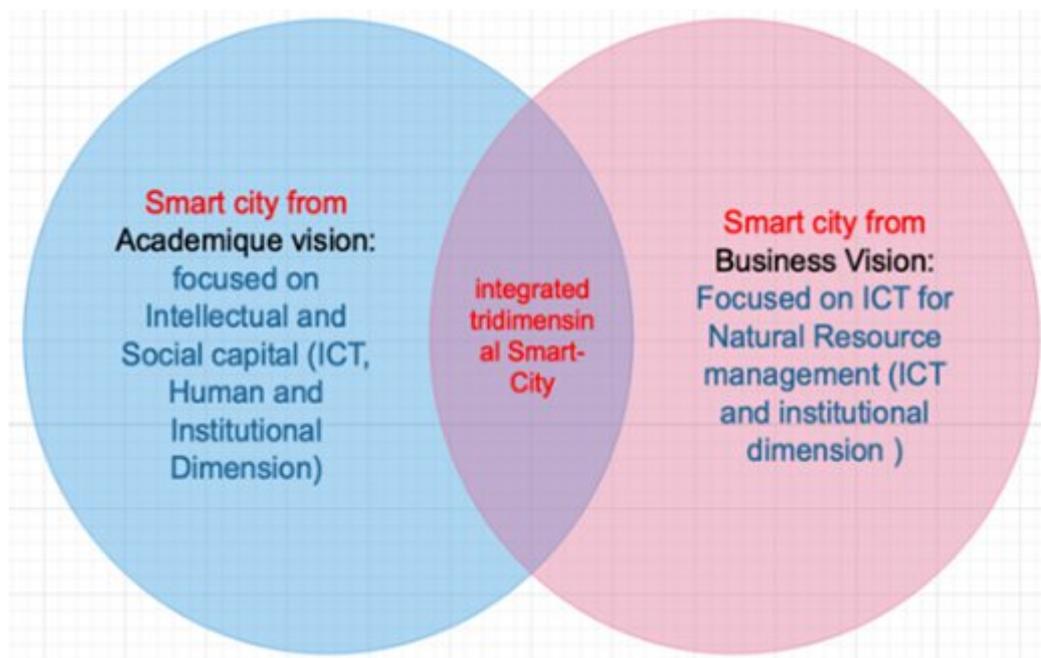

(c) Dameri's Smart City Conceptual Model [Dameri, 2017]

Figure 1. Smart City Models. (A) Cohen's Smart City Wheel. (B) Pardo's Smart City Model. (C) Dameri's Smart City Model.

Though there is no generic or universally agreed definition of the "Smart City" concept, it is important to emphasize that some authors have focused on defining the human, social, cultural, and economic dimensions and the role they play. Bibri [2018a] focused on technologies, whose role is to improve the economy, and society and ensure environmental sustainability. Neirotti et al. [2014] focused on the efforts made to improve the quality of life of citizens. Dameri [2017] considered two types of visions when presenting the dimensions of SC, these are the "Academic Vision" and the "Business Vision". For this author, the academic vision favors two capitals that directly affect the citizens of a city: intellectual capital and social or welfare capital. The author described a SC as a "city of knowledge" hence, focused on the education of the citizenry and their social well-being while relying on digital infrastructure. On the other hand, the "Business vision" is considered three-dimensional, encompassing human, technological, and institutional capital. Emphasis was placed on technology, especially ICT, which the author considered to be the main element in environmental management. Finally, the Academic vision advocates for the resolution of problems linked to pollution, traffic, waste management and treatment, reduction of energy demand and consumption, and quality of water and air among others; while the Business Vision is driven by ICT and acts on the components of a city including the land, the government, the infrastructure and the people [Dameri, 2017].

From these articles, we can surmise that the dimensions of SC are paramount for its definition. None of these frameworks can be considered better than the other; as each justifies its orientation, however, Cohen's model [Cohen, 2012, Cohen 2014] seems more theoretically detailed and was also applied in Giffinger et al. [2007], Benamrou et al. [2016]. Of these frameworks, those of Cohen [2012], Nam and Pardo [2011], and Dameri [2017] stand out, as they are the most widely adopted and are depicted in Fig. 1. We therefore strategically focus on these three and deduce a definition for SC from their point of convergence.

Though dimensions are vital, context still has to be taken into account. From Dameri [2017] a smart city is considered "a smart community of smart people". This implies that in defining SC, the intelligence of the citizenry is given higher priority than the social or welfare vision. In this context, a SC is, firstly, a city of knowledge and a high level of education; then a city of social well-being of its population, based on digital transformation. In this vein, Europe for instance can be considered leaning towards Cohen's model [Giffinger et al., 2007, Benamrou et al., 2016, Cohen]. This knowledge-first view seems like an agreeable definition as it relates to our lives today. Today, the SC concept is being widely researched in universities across the world and is the subject of many research publications on urban development. It can easily be argued that there can be no real urban development (or political glory) without an extensive scientific study on the subject. This is evident in many industrial and government projects around us today, which go through extensive research, simulations, and feasibility studies before actual deployment. This view is also buttressed in literature [Dameri, 2017, Bibri, 2018a,b, 2019].

## 2.2 Smart City Related Concepts

### 2.2.1 Urban Dynamics

The management and security of inhabitants is paramount in SC. According to the UN, the World's population will grow by 4 billion and urbanization will reach 70% in the coming years [UN, 2023, MLIT, 2019]. The increase or decrease of the population in a city can have enormous impacts on its demographics - aging, mortality, and natality rates. Houses, buildings, industries, schools, and other infrastructures are built (or disappear) according to constraints and necessities that emerge and interact with them. These make SC dynamic in nature. Two articles that have influenced the planning of urban capacities, shapes, structures, resources, and energy management are: i.) the works of Wolman [1969], where a model and tools for understanding and structuring "Urban dynamics" was proposed, and ii.) Alfeld [1995], who suggested that the institutionalization of "systems dynamics" can be a vital foundation for the formation of a new generation of urban leaders.

By applying system dynamics in studying the development of cities, it has been shown that it is possible to bring lasting change to a city as well as balance to its structures. Urban dynamics is therefore a corollary of SC. The different definitions, frameworks (models), and initiatives of SC reveal that it is a dynamic and complex system. A city that is transformed by digitization would have a digital nervous system consisting of system dynamics. Such system dynamics are ICT-powered, which in themselves are dynamic and constantly evolving. Weiser [1991] had envisaged that the best technologies are the invisible ones. In the vision of future SC, the conceptual connection between "Smart cities" and "Urban dynamics" would be inevitable and predominantly invisible. This can be seen, for example, in the field of Ecology and Agriculture [MILT, 2019, Duran-Encalada and Pauca-Careres, 2009], where the underlying technology powering smart ecology or agriculture is completely transparent.

### 2.2.2 Sustainability

The concept of "Sustainability" goes hand in hand with that of the dynamic development of a city. As with SC, there are diverse definitions of a sustainable city or what one might look like [Roseland, 2001, Jabareen, 2006, Waas et al., 2010, Register, 2013, Rabari and Storper, 2015, Martos et al., 2016, Bibri and Krogstie, 2017, Dwevedi et al., 2018]. There is no canonical definition of the concept of "Sustainability". One of the least ambiguous definitions is that of Bibri and Krogstie [2017], which defines a sustainable city as any city where social life, promotion of the preservation of the environment against pollution and poor waste management, efficient use of renewable energies, significant promotion of ICT applications and innovations, reduced private transportation for the benefit of public transport and walking, good management, sharing of housing spaces and economic resources, and good access to education and health, etc. are paramount. This definition incorporates elements of smartness, hence sustainable cities are inherently smart and are often referred to as Smart Sustainable City.

New sets of technological phenomena come into play with regard to Sustainable Cities. It is the pervasive and massive use of advanced ICT to interconnect urban areas & activities, share, analyze & synthesize data, and communicate through infrastructures, complex systems of networks, services, machines, and individuals [Bibri and Krogstie, 2017]. Indeed, the sustainable development of cities and societies can only be achieved at the intersection between the development of science and technology. From this, applications where ambient intelligence is combined with the Internet of Things (IoT) for domestic and industrial use, or in transport, environment, health, and other services, can emerge. Again the underlying technologies, which support/power these services have to remain transparent to the users [Weiser, 1991].

The authors in Martos et al. [2016], argued that the sustainability of urban development is based on four dimensions, which are: shape, environment, economy, and equity. The shape of the city is the most important as it affects the other dimensions. It can be viewed from four spatial levels: the regional level, the city level, the community level, and the building level. According to Jabareen [2006] in [Pichler, 2017], analysis of these spatial levels can be done using seven key concepts: compactness, transportation, density, mixed land use, diversity, passive solar design, and greening. City shape aside, a sustainable city, is also defined by the presence of sustainable urban transportation, energy-conserving buildings, urban green areas, municipal solid waste management & recycling systems, water supply, and resilience to social variables [Bibri, 2018b].

### 2.2.3 Urban ICT and Computing

Currently, we are in an era of the "digital skin of cities" where sensors measure, detect, and collect diverse data from the urban environment and transport the same through ICT. In these new cities, IoT and sensor networks are used to obtain Big data on sectors, domains, activities, services, inhabitants, events, and other factors related to a city to improve the well-being and quality of life of the citizens. The data can then be shared, searched, classified, analyzed, secured, visualized, and/or stored. In essence, the SC of today is one in which ICT regulates the economy, the quality of life, security, power, transportation, and education [Rabari and Storper, 2015, Martos et al., 2016, Dwevedi et al., 2018].

To achieve this, a sustainable connection fabric (link) between urban ICT / computing and cities has to be woven. Sustainability and robustness are paramount for this link, as the management and governance of these ultramodern and data-driven cities would be built on it. It would interconnect numerous digitally-enabled objects, network nodes, devices with control posts, regulators, billing agencies, and government for monitoring and analysis. There would also be a variety of participants including, IoT compatible machines, autonomous vehicles, private individuals, private and public organizations, as well as governmental parastatals. Machine learning, Data science, and analytical frameworks would be pivotal in this urban ICT and data-driven society [Roseland, 2001, Register, 2013, Dameri, 2017, Ajayi et al., 2022].

**2.2.4 Eco-city Concept**

Since the early 1970s, there have been growing calls for ecologically aware cities. These calls focus on climate change, care for the environment, and sustainable development. Several related works on ecology-friendly/conscious cities or 'Eco-cities' have been surveyed by Roseland [1997].

On one hand, Eco-cities advocate for sustainable urban innovations and clean environments. They are concerned with environmental management, including waste management, atmospheric pollution, renewable energy, green spaces, water resources, climate change, land use, geography, and meteorology. Many of these requirements can be achieved using ICT (IoT and Big data analytics). On the other hand, in earlier sections, we have established that a SC is one that leverages on ICT for many of its operations. Thus, ICT can be a bridge that links Eco-cities with Smart cities. This implies that by relying on ICT a SC can incorporate ecological awareness into one of its dimensions; hence, a SC can be considered an Eco-city. Several work have shown this inter-play, such as in Roseland [1997, 2001], Berthold and Hoglund Wetterwik [2013], where the authors discussed the multi-dimensionality of Eco- and Smart cities, as well as Register [2013] who discussed the guiding principles for transforming or building an ecologically conscious SC.

## 3. SMART VILLAGE

Alongside the concept of Smart City (SC), the emergence of Smart Village (SV) as a complementary concept has been transformative, especially in developing nations. While the USA, European countries, and other developed regions have harnessed the potential of Smart Cities, developing countries, with India at the forefront, have recognized the immense promise of Smart Villages [Varghese, 2016a]. In these regions, it's widely understood that the true impact of Smart Cities cannot be realized without concurrent development in rural areas [Viswanadham and Vedula, 2010]. The concept of Smart Village began to take shape around 2010, aiming to create a comprehensive ecosystem for rural areas (ibid.). It was conceived as a solution to enhance the quality of life in rural regions while addressing pressing issues like urban overpopulation. Projections indicate that by 2030, accelerated urbanization will result in overcrowded cities globally. This prediction aligns with the United Nations' forecast of the global population reaching approximately 10 billion by 2050 [UN DESA, 2022]. Urbanization undeniably exerts a magnetic pull on people, yet the smarter a city becomes, the more it experiences population growth and associated challenges. In many developing countries, a rural exodus is looming, with villages often seen as lacking in economic prospects. Young people actively migrate from villages to cities, depleting the economic vitality of rural areas while increasing the pressure on urban centers [Fajrillah et al., 2018]. This migration, coupled with population growth, contributes to global carbon emissions and introduces technological challenges, including strained infrastructure and security concerns. Fig. 2 gives a high-level pictorial illustration of our SV model and shows the essential building blocks of an SV.

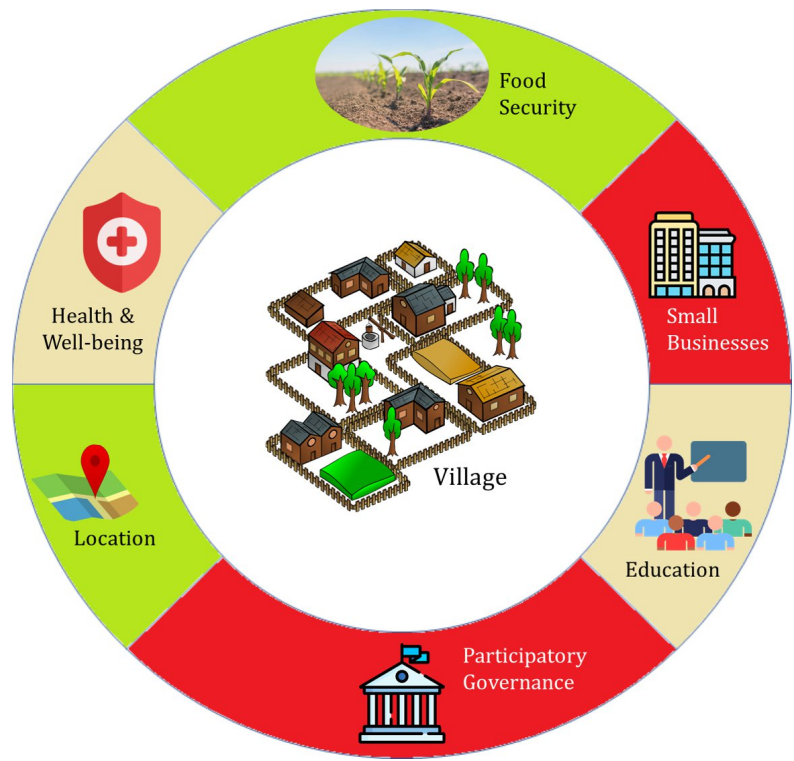

Figure 2. Services within smart villages

Smart Village initiatives offer compelling solutions to the challenges outlined above and have the potential to profoundly impact the socio-economic and cultural aspects of rural life. These initiatives can serve as:

• Climate Change Mitigators: By embracing sustainable practices and green technologies, SVs have the potential to act as stabilization solutions to climate change. These initiatives reduce the carbon footprint and promote eco-friendly practices, contributing to global environmental sustainability.

• Hubs for Intelligent Agriculture: SVs can transform into sources of intelligent agricultural production and livestock breeding. The incorporation of modern farming techniques, data-driven agriculture, and precision farming can significantly enhance food security, economic viability, and employment opportunities in rural areas.

• Communities of Quality Living: SVs have the potential to foster networks of communities that provide access to a high quality of life. Initiatives can improve access to healthcare, education, and essential services, leading to an improved standard of living for rural residents.

• Educational Hubs: By facilitating education and awareness, SV initiatives can bridge the knowledge gap. Access to fundamental education and digital resources can empower rural communities, providing them with the tools needed for personal development and awareness.

• Economic Development Engines: Local businesses and entrepreneurship thrive in the SV model. These initiatives can stimulate economic development by creating jobs, encouraging small-scale enterprises, and supporting local commerce.

• Cultural Preservation and Enhancement: SV initiatives can empower local cultures and traditions. By promoting cultural preservation and revival, these initiatives can enhance the cultural identity of rural communities.

It is important to note that the framework of SV initiatives may vary across regions of the world, reflecting the unique needs and contexts of each community. Geographical information, therefore, plays a pivotal role in SV projects [Azziza and Susanto, 2019]. Examples of Key Smart Village Initiatives:

• European Digitization Initiatives: Zavratnik et al. [2020] discussed digitization efforts in European rural communities, showcasing how technology can positively transform these areas, promote economic growth, and bridge the urban-rural digital divide.

• IoT Applications in Indian Rural Areas: Natarajan and Kumar [2017] explored the application of the Internet of Things (IoT) to enhance the quality of life, agriculture, and healthcare in rural regions of India, with a focus on enabling technology-driven rural development.

• Global SV Initiatives: SV projects have been launched in developing countries across East and West Africa, South Asia, South America, and Central America, demonstrating the global reach and impact of SV initiatives [Zavratnik et al., 2020].

As highlighted above, the potential socioeconomic and cultural impact of these 'SV initiatives provides concrete evidence of the positive transformations that SVs can bring to various regions of the world.

## 4. SMART CITY PROJECTS AND GLOBAL INITIATIVES

Smart cities are global initiatives, most heavily dependent on the application of real-time ICT, the ubiquity of computers, network connectivity, and integrated services. Albino et al. [2015] stated that smart city initiatives in several countries across Europe, the USA, and Asia are considered national urban development projects. These projects epitomize the significance and role of advanced ICT, especially Big data analytics, in enhancing the operations, functions, services, strategies, and policies of SCs of the future; with regards to planning, management, development, and governance. In this section we briefly discuss a few SC initiatives globally and their impacts.

Taken positively, this paper would perhaps be too long if all data on the practical impact of SC initiatives gathered were to be discussed in this survey. Out of approximately 152 projects identified globally, there were 35 smart city initiative projects in North America, 47 in Europe, 50 in Asia, 10 in South America, and 10 in Africa. From these numbers, we can infer that comparatively there is less data online on SC initiatives in South America and Africa than in other parts of the world. This might be attributed to several reasons, not limited to poor (or no) documentation, limited accessibility to Internet-based information repositories, and the absence of (or little) actual SC deployments in these regions.

In 2010, the European Commission (EC) launched several smart city initiative projects that were to be completed in 2020. Though the completion of many of these projects was delayed due to the global COVID-19 pandemic, most are in their final phases. These initiatives were based on the "Smart City Wheel" model defined by Cohen [2012] and shown in Fig. 1a. Using the work of Alaverdyen et al. [2018] as a reference, the implementation of the EC's SC initiatives can be grouped into six dimensions, paired as follows:

• Smart Environment and Smart Mobility: initiatives focusing on climate change and energy.

• Smart Living and Smart Governance: initiatives aimed at improving the quality of life of the citizenry and the fight against poverty and social exclusion.

• Smart Economy and Smart People: initiatives geared toward education and job creation.

These initiatives were part of the sustainable development programs for the EC and leveraged several ICT applications centered on Big Data, 5G, and the IoT. The implementation policy is based on tools for raising awareness, promoting understanding, collaboration, and participation of institutionalized actors. The cities where these initiatives were to be deployed are grouped into 3 categories: Smart City Clusters, Living Labs, and Best Practices Cases. 11 countries were selected as SC Clusters, namely: Belgium, Cyprus, Czech Republic, Estonia, Finland, France, Italy, Romania, Slovakia, Spain, Sweden. Living Labs were created in Amsterdam, Barcelona, and Helsinki, while Amsterdam, Barcelona, Copenhagen, Dublin, Helsinki, and Manchester are examples of Best Practices Cases. In North America, cities such as Seattle, Quebec, and Ottawa are examples of SC; while in South America, Brazil has a considerable number of SC including Sao Paulo, Vitoria, Brasilia, and Rio Je Janeiro.

From our findings and as at the time of writing, Asia had the highest number of SC initiatives. However, it must be noted that these were not evenly spread across the Asian continents but rather concentrated in the Far East.

Based on Cohen's diagram, deductions by Gantori [2019] in [UBS, 2019], clearly show that Asia is growing in initiatives. China, Japan, Singapore, Hong Kong, and India are leading the charge for SC, with several digitization efforts already deployed. In China, for example, the technological pillars on which most smart city projects are built are IoT, Big Data, Cloud Computing, and other smart systems. Coincidentally, these are also the enabling technologies of the fourth industrial revolution (4IR) [Ajayi et al., 2022]. It can therefore be argued that smart cities are a direct actualization of the Asian 4IR. As of 2017, there were about 500 SC initiatives in China at various stages of development Gantori [2019]. It is predicted that by 2025, Asia will experience 13% penetration of smart connectivity, 10% smart automation, 9% of smart mobility, 30% of smart services, 15% of smart healthcare, and 23% of smart governance UBS [2019]. With the aggressive rate at which China is adapting the SC concept, many have tipped China to become the world leader in smart cities in the foreseeable future CAICT and MIIT [2014].

Africa has its own SC initiatives, though only a few are known, possibly as a result of poor documentation. Albino et al. [2015] identified 10 African SC initiatives, with the North African countries leading the deployment count. Morocco is considered to be the pioneer of the African SC. In Algeria, a SC experience was initiated in 2011 with the new city of Sidi Abdellah, a city located west of Algiers. As at the time of writing, the Sidi Abdellah project was still incomplete – perhaps due to disruptions from the Arab Spring unrest. The "Algiers Smart City" project, planned for 2035 in Algeria, is another notable example of SC in Africa. Algeria has developed a working model for the SC project through a strategic collaboration between start-up companies, research and development laboratories, large investors, and Universities. The diversity in stakeholders ensures that the SC plans are holistic and take into account the comparison and analysis of failures and successes of previous experiences for future smart cities [Ait-Yahia et al. 2019]. According to Aıt-Yahia et al. [2019], Tunisia aims to make Bizerte a smart city by 2050 . In East Africa, Rwanda has taken notable strides towards a sustainable city and is commonly referred to as the "Singapore of Africa". The rapid economic development of Rwanda can be attributed to the emphasis placed on Education, hence, we can conclude that the Dameri model is prevalent in the country.

The 2014 UN study on World Urbanization Prospects shows that about 40% of the Sub-Saharan African population lives in urban environments [UN, 2014]. That report, as well as Lee [2014], show several factors that make African cities prime candidates for SC evolution. Notable factors include minimal legacy disadvantages, a youthful population, and rapid urbanization. Taking these identified factors and the demographic profile of African cities (large cities with thousands of inhabitants) into account, a number of countries perfectly fit the bill for African SC initiatives. These include Nigeria, South Africa, Kenya, Morocco, Algeria, Egypt, Kenya and the Democratic Republic of Congo (DRC). The specific viability of cities in most of these countries has already been discussed in [Giles, 2017]. Though poorly documented, a number of these countries have taken (or are taking) concrete steps to convert major cities into SCs based on one or more dimensions of the Cohen model as reported in Duran-Encalada and Paucar-Careres [2009], Nkurikiyimfura [2016], Odendaal [2016], Mekni and Huard [2020].

# 5. COMPARISON OF SMART CITY INITIATIVES IN EUROPE AND AFRICA

Different models have been adapted by various SC initiatives globally. For instance, the Cohen model is the most used in European, but this is not the case elsewhere in the world, particularly in Africa and Asia. Irrespective of the model adopted, the points of convergence appear to be those that are brought together in the three models we presented above – the Dameri model, the Prado model, and the Cohen model (Fig. 1). In this section, we briefly survey notable SC initiatives in Europe and Africa.

## 5.1 Survey of European Smart Cities Initiatives

With regards to smart city initiatives in Europe, we reviewed a selection of 37 cities across 13 European countries in 2014 and summarized these on Table 1. This table is based on the 2014 smart city mapping study carried out by the Directorate General for Internal Policies [Waas et al., 2010]. Six characteristics based on the Cohen model were used as a benchmark, viz.: Smart Governance, Economy, Mobility, Environment, People and Living. From the table, only 6 of the 37 cities were able to fulfill 100% of a chosen characteristic, namely Amsterdam, Barcelona, Copenhagen, Dublin, Helsinki, and Manchester. Furthermore, Smart Environment was the most implemented, with 36 of the 37 cities implementing it. Smart Government was implemented in 19 cities, Smart Economy in 18 cities, Smart Mobility in 20; Smart People in 26, and Smart Living in 20 cities. From a country perspective, as of 2014, Germany had the most SC initiatives with 8 projects. It is followed by Denmark (6), and the Netherlands, Spain, and the UK with 5 each. Finland had 4 while Sweden and Austria had 3 and 2 respectively.

Table 1: Selected Smart Cities in Europe [Manville et al., 2014]

| City | Country | Project Count | Smart Govt. | Smart Economy | Smart Mobility | Smart Environs | Smart People | Smart Living |
|---|---|---|---|---|---|---|---|---|
| Vienna | Austria | 2 | | | ✓ | ✓ | ✓ | ✓ |
| Antwerp | Belgium | 1 | | | | ✓ | ✓ | ✓ |
| Copenhagen | Denmark | 5 | ✓ | ✓ | ✓ | ✓ | ✓ | ✓ |
| Aartus | Denmark | 1 | | ✓ | | ✓ | ✓ | |
| Tallinn | Estonia | 1 | | ✓ | | ✓ | ✓ | |
| Helsinki | Finland | 2 | ✓ | ✓ | ✓ | ✓ | ✓ | ✓ |
| Oulu | Finland | 1 | ✓ | ✓ | | ✓ | ✓ | ✓ |
| Tampere | Finland | 1 | ✓ | ✓ | | ✓ | ✓ | |
| Lyon | France | 1 | | | | ✓ | | ✓ |
| Bremen | Germany | 1 | | | ✓ | ✓ | ✓ | ✓ |
| Cologne | Germany | 3 | ✓ | ✓ | | ✓ | ✓ | ✓ |
| Hamburg | Germany | 2 | | ✓ | ✓ | ✓ | ✓ | ✓ |
| Mannhien | Germany | 1 | | ✓ | | ✓ | ✓ | |
| Munich | Germany | 1 | ✓ | | | ✓ | ✓ | |
| Athens | Greece | 1 | ✓ | | | ✓ | ✓ | ✓ |
| Thessaloniki | Greece | 1 | ✓ | ✓ | ✓ | ✓ | | |
| Budapest | Hungary | 1 | | | | ✓ | ✓ | ✓ |
| Miskolc | Hungary | 1 | | ✓ | | ✓ | | |
| Dublin | Ireland | 1 | ✓ | ✓ | ✓ | ✓ | ✓ | ✓ |
| Florence | Italy | 1 | ✓ | | | ✓ | | |
| Milan | Italy | 1 | | ✓ | | ✓ | ✓ | |
| Amsterdam | Netherlands | 2 | ✓ | ✓ | ✓ | ✓ | ✓ | ✓ |
| Eindhoven | Netherlands | 1 | | | ✓ | ✓ | ✓ | |
| Enschede | Netherlands | 1 | | | ✓ | | | |

| City | Country | Project Count | Smart Govt. | Smart Economy | Smart Mobility | Smart Environs | Smart People | Smart Living |
|---|---|---|---|---|---|---|---|---|
| Tilburg | Netherlands | 1 | | | | ✓ | ✓ | |
| Tirgu Mures | Romania | 1 | ✓ | | | ✓ | ✓ | ✓ |
| Ljubijana | Slovenia | 1 | | | ✓ | ✓ | | ✓ |
| Barcelona | Spain | 3 | ✓ | ✓ | ✓ | ✓ | ✓ | ✓ |
| Bilbao | Spain | 1 | | ✓ | ✓ | ✓ | | ✓ |
| Zaragoza | Spain | 1 | | | ✓ | ✓ | | |
| Gothenburg | Sweden | 1 | ✓ | | | ✓ | | |
| Malmo | Sweden | 1 | ✓ | ✓ | | ✓ | ✓ | ✓ |
| Stockholm | Sweden | 1 | ✓ | | ✓ | ✓ | ✓ | ✓ |
| Coventry | UK | 1 | ✓ | | | ✓ | | |
| Glasgow | UK | 1 | | | ✓ | ✓ | ✓ | ✓ |
| London | UK | 2 | ✓ | ✓ | ✓ | ✓ | ✓ | |
| Manchester | UK | 1 | ✓ | ✓ | ✓ | ✓ | ✓ | ✓ |

**5.2 Survey of African Smart Cities Initiatives**

Unfortunately, Africa has not followed this trend in Europe. Projects in Africa are few and widely dispersed. Within the same time frame as the first European report (2013-2015), the Casablanca Smart City cluster in Morocco was the most notable example and was launched in 2013 [Cohen, 2014]. Within this cluster, the "Ville Verte Mohamed VI" is a Smart Economy and Smart Environment [Pieterse and Zevi, 2018] project. Rwanda has the most SC projects, especially in Kigali. The country has implemented three forms of Cohen model-oriented projects divided into 9 building blocks and split into 27 smart-city initiatives. These are as depicted in "SC Rwanda Action Plan Version 2 Master plan Version 2.0" [Rich et al., 2017]. This Rwandan vision suggests that SC initiatives should focus on two key areas - City Flow and City Services. The Rwandan template can thus be adapted as a guide for African smart city initiatives. Despite not finding unique assessment reports of SCs in Africa similar to those in Europe [Manville et al., 2014], the information we found in the literature is summarized in Table 2. The table also shows our classification of known African smart city initiatives based on Cohen's model.

Table 2: African Smart City initiatives based on Boyd Cohen's mode

| City | Country | Project Count | Smart Govt. | Smart Economy | Smart Mobility | Smart Environs | Smart People | Smart Living | Ref. |
|---|---|---|---|---|---|---|---|---|---|
| Kigali | Rwanda | 16 | ✓ | ✓ | ✓ | ✓ | ✓ | ✓ | Nkurikiyimfura 2016, Rich et al. 2017 |
| Nairobi | Kenya | 2 | ✓ | ✓ | ✓ | ✓ | ✓ | ✓ | Fernandez-Anes et al., 2018, Lamari & Oukarfi, 2018, Smartcity 2018 |
| Casablanca | Morocco | 4 | | ✓ | ✓ | ✓ | ✓ | ✓ | Benamrou, 2016, Angelidou, 2017 |
| Cape Town | South Africa | 5 | ✓ | ✓ | ✓ | ✓ | ✓ | ✓ | Smartcity 2018, Manirakiza et al., 2019, Bayu, 2020 |

Though the Rwandan Smart City Master plan [Rich et al., 2017] indicates 27 smart city initiatives, we could only find 16. Like with the Rwandan projects, details of each smart city initiative in Africa, such as scope, size, and cost factors were not readily available. However, a high-level comparative study of the magnitude of each project reveals wide dispersion in scale across the various projects. For example, the four Casablanca SC initiatives in Morocco, which are the Social Sustainable Solar Smart City, Virtual Museum of Casablanca, the development of urban Video surveillance, and the Casablanca digital project [Benamrou et al., 2016] dwarf Kenya's Konza Technology City, Garden City, Smart Nairobi, and Intelligent Satellite Cities (consisting of Tatu City and Machakos New City) projects [Maslon-Oracz and Mazurewicz, 2015]. Beyond these, several other SC initiatives have sprung up across Africa in the last decade, such as King City in Ghana, Eko Atlantic City in Lagos, Nigeria, and Waterfall City in South Africa. Most of which are collaborative governance between the private and public sectors. As of the time of writing, many of these projects are still in their first few phases, and unfortunately not well documented.

**5.3. Transition from Smart City to Smart Village**

When examining the Smart City (SC) projects that have been initiated in African cities, it becomes evident that many of them lack a clear vision of "Smart growth." Several African countries tend to identify certain characteristics of a Smart City without first considering the concept of Smart Villages. This is particularly concerning given that a majority of the population in these countries resides in villages and towns. In essence, the actual transition from Smart Villages to Smart Cities, which we refer to as "smart growth," is often neglected.

Smart Villages are emerging as satellite smart cities in their own right. Therefore, it is important to recognize the complementary nature of Smart Cities and Smart Villages. Considering that information and communication technology (ICT) flows and services related to human-environment interaction are synonymous across both Smart Cities and Smart Villages, it is feasible to develop an "SC-SV Ontology" catalog. This ontology would facilitate the easy adaptation and reuse of Smart City templates and frameworks, as well as enable interoperability and interconnection between both concepts. Such a catalog would significantly streamline the transition from smart villages to smart cities across Africa and potentially contribute to the United Nations' acceleration of the achievement of Sustainable Development Goals (SDGs) in Africa and other developing countries globally.

It is important to note that Smart City initiatives vary significantly and do not produce uniform impacts or address the same challenges. While we have identified ICT as a potential baseline, certain prerequisites must be in place, particularly in developing countries. For example, one essential prerequisite is access to electricity. Moreover, it is crucial to consider the human dimension in Smart City development, as cities are ultimately built for human occupancy. However, it's worth noting that population growth typically follows the establishment of a Smart City. Therefore, the initial focus should be on incorporating the capacity for future population growth rather than solely on the population at the initialization stage.

Research plays a fundamental role in the development of a Smart City. Prior to launching an SC initiative, comprehensive research is essential. Such research should address critical questions related to the objective, scope, target audience, stakeholders, methodology, costs, and associated risks of these projects. The more comprehensive the research conducted, the richer the "SC ontology" will be in terms of human experiences, interactions between humans and machines, the environment, and the various subsystems of the Smart City [Komninos et al., 2019, Gyrard et al., 2018]. Therefore, research serves as the ideal starting point for an SC initiative, providing valuable insights and data that can inform the design and development of a Smart City.

**6. THE LUBUMBASHI CASE STUDY**

The city of Lubumbashi is the economic capital of DRC and is the second largest and second-most populous city in the country. Despite meeting several criteria that qualify it as a suitable candidate for SC, it is not yet a SC. Moreover, at the time of writing, there is no SC in DRC. Lubumbashi has been in existence since 1910 and is located within geographical coordinates of 11°40′11″S, 27°29′11″E, sitting at an altitude of

1208m. Statistics from the National Institute of Statistics [CityPopulation, 2020] show that the city has a surface area of 747 $Km^2$ and an estimate of between 1,200,000 and 1,800,000 inhabitants as of the year 2020. This statistic translates to an approximate population density of 4,000 inhabitants/$Km^2$ [Wikipedia, 2020].

Though theoretically Lubumbashi is an ideal SC candidate, there are several limiting fundamental factors that need to be addressed before it can transition to an SC. These include but are not limited to addressing the primary needs, such as access to portable drinking water, good road networks, electricity, and the Internet. As an example, the poor road network, limited formal house numbering system and lack of coordinated maps make locating addresses in certain parts of the city very difficult. Coupling this with limited or no electricity and the Internet makes it next to impossible to successfully deploy e-commerce or courier delivery services in the city. Similar challenges are prevalent in numerous other cities across sub-Saharan Africa [Lee, 2014].

Table 3: Some African and European Smart Cities with similar population as Lubumbashi in 2020

| City | Country | Status | Approx. Population |
|---|---|---|---|
| Casablanca | Morocco | SC initiatives underway | 4,475,000 |
| Alexandria | Egypt | SC initiatives underway | 5,800,000 |
| Nairobi | Kenya | SC initiatives underway | 5,900,000 |
| Cape Town | South Africa | SC initiatives underway | 4,225,000 |
| Kigali | Rwanda | SC initiatives underway | 1,132,000 |
| Lubumbashi | DR Congo | Nil | 1,600,000 |
| Amsterdam | Netherlands | SC initiatives underway | 2,475,000 |
| Manchester | Great Britain | SC initiatives underway | 3,050,000 |
| Barcelona | Spain | SC initiatives underway | 4,775,000 |

Table 3 shows a number of African cities where SC initiatives have been deployed and have been transformed into (or have made notable strides in becoming) smart cities in a space of 3-5 years. We also included a few cities in Europe for comparative purposes. Lubumbashi City can adapt templates and experiences from some of these cities to plan its own SC initiative project. To achieve this, the city could start with a smart growth logic, beginning with a transformation of small villages/communities to smart villages, and then to smart cities. Base parameters need to be defined, from which either a reconstruction or upgrade of existing infrastructure would be done or new living labs would be constructed.

From a literary standpoint, to build a comparative table of SC publications across a few cities in Africa, using a simple methodology wherein we:

1. randomly selected three of the 9 cities compared on Table 3: Casablanca, Nairobi, and Kigali.
2. formulated a search phrase in the following way: "name of the city" followed by the term "SC initiative". Example "Lubumbashi SC initiative".
3. ran our query on Google Scholar.
4. specified a period from 2016 to 2020 and restricted data to results on the first page only.

Data collected are summarized in Table 4.

Table 4: Scientific publications on SC in Casablanca, Nairobi, Kigali, and Lubumbashi

| References | City | Status |
|---|---|---|
| Benamrou, 2016, Khomsi & Bedard, 2016, Laaboudi, 2016, Lamari & Oukarfi, 2018, Fernandez-Anes et al., 2018, El Haj and Ait, 2020 | Casablanca | 4 major SCs initiatives |
| Duran-Encalada & Paucar-Careres, 2009, | Nairobi | 2 major SCs and 3 satellite |

| Nkurikiyimfura, 2016, Angelidou, 2017, Fernandez-Anes et al., 2018, Lamari & Oukarfi, 2018] | | initiatives |
| --- | --- | --- |
| Maslon-Oracz and M. Mazurewicz, 2015, Rich et al., 2017, Manirakiza et al., 2019, Bayu, 2020 | Kigali | Several SC initiatives |
| Nil | Lubumbashi | None |

To succeed, it is imperative that prospective African SC adhere to models of successfully deployed African SC initiatives or those from Asia, rather than blindly adopting models from the Western world. Africa is undergoing rapid urbanization Pieterse and Zevi [2018], with a young population growing at an annual rate of 3.5%. Statistical studies suggest that about 50% of Africans will be living in urban environments by 2030. While there is, in general, a correlation and causality between urbanization and the economic growth of a country and its cities, the rapid urbanization in Africa is quite different and this might pose a problem. Africa's growth is not a direct consequence of industrialization, hence there is no corresponding infrastructural growth, but rather from illiteracy and poverty. There is also the mass exodus of youth to cities, which inevitably leads to the overpopulation of such cities. Without proper measures in place, the population of the reference SCs in Africa (shown in Table 3) might result in challenges stemming from urbanization [Bayu, 2020].

Realizing the UN's SDGs in developing worlds would require a comprehensive knowledge of the country/region, regional peculiarities (culture, policies, etc.), constituent entities, and the interrelationships between the entities and their environment. It would be useful to create or revisit the smart city ontologies for this reason. As stated in section 4 above, knowledge about the basic obstacles, priorities, population growth, and prospective sustainable targets is the first step to any SC project. In our use case city of Lubumbashi, a prominent challenge being faced is access to basic resources and utilities. This challenge is significant and greatly hampers the city's ability to transition to SCs and not get left behind in the global march towards the 4IR (Fourth Industrial Revolution). Beyond Lubumbashi, the economic power and levels of infrastructural development in many sub-Saharan African cities are still poor to support the 4IR. Though collaborative efforts, such as those proposed in [Ajayi et al., 2019, 2020], could be a viable solution, such solutions are at a much higher level. Smart village initiatives can thus be pivots for achieving economic growth at the grassroots level. This is also buttressed in the Africa 2063 report Pieterse and Zevi [2018]. A strong understanding of the conceptualization and contextualization of smart city projects through the SC ontologies will make real African smart cities.

For the city of Lubumbashi's SC initiative, we make the following recommendations:

1. the city may draw inspiration from Morocco's Casablanca SC project and use the intelligence characteristics and indicators in Giffinger et al. [2007], Benamrou et al. [2016] as guidelines.

2. The SC initiative should start off with the Dameri model [Dameri, 2017] to identify the baseline requirements and constraints, then implement the Cohen model [Cohen, 2012]. This model with its 6 characteristics, 18 variables, and 62 indicators, has been applied in SC initiatives in over 30 cities across Europe, North America, Latin America, and the Asia Pacific.

3. SC evaluation be based on the z-score standardization used by Giffinger et al. [2007]. Morocco and other African cities have used this evaluation system directly or indirectly to assess the final features of their smart city initiatives.

## 7. CONCLUSIONS

Smart Cities (SCs) are beneficial and necessary for the development of a country and the well-being of the people who live there. However, there is yet to be a unified definition for the Smart city concept. This work sought to find this definition by exploring literature to determine the constituent components of SCs as well as the fabric interconnecting them. For smart growth to occur, particularly in developing nations, the concept of a Smart city must be predicated on the development of Smart Villages (SVs). This is because the majority of residents in developing countries dwell in villages, hence concentrating on creating SCs in developing countries would be lopsided and counterproductive if the villages are forgotten. A means by which developing nations can catch up with the Western world is by replicating models that worked in these developed countries. However, there is no single standardized model that can be adopted for all instances, rather implemented model(s) must first take into account the local context, the peculiarities therein, and the fundamental requirements. This paper considered three SC models with a view of identifying which could be well suited for developing countries. After reviewing some SC initiatives through the lenses of the 3 common SC models - Cohen, Pardo, and Dameri models, we infer that African countries, like other developing countries globally, must consider education as the first and primary take-off point when considering SC or SV projects. Once the intellectual capital has been established, bespoke SC/SV initiatives can then be implemented, which address the peculiarities of the city being considered. The third step would then be to benchmark the SC/SV initiative using the established models. In the context of African SC initiatives, our survey revealed that literature on African SC/SV is scarce, possibly due to poor documentation of the project or limited access to Internet-based repositories. Finally, the city of Lubumbashi in the DR Congo was considered as a use-case for deploying SC/SV in Africa. The unique features of the city were explored and guidelines for converting Lubumbashi to an SC were presented. Notable among which is leveraging on experience SCs deployed in neighboring countries such as Nairobi in Kenya and Kigali in Rwanda.

In the future, a comprehensive survey on the status of all SC initiatives across Africa could be considered. Furthermore, developing a dynamic model for smart education or smart knowledge projects that can be replicated across African cities might be considered another avenue for expanding this work. Finally, building an ontological model of growth from Smart villages to Smart Cities might also be another interesting direction by which this work can be expanded in future works.

# References

K.G. Aıt-Yahia, Faouzi Ghidouche, and Gilles N'Goala. Smart city of algiers: defining its context. In Smart city emergence, pages 391–405. Elsevier, 2019.

O. Ajayi, A. Bagula, and K. Ma. Fourth industrial revolution for development: The relevance of cloud federation in healthcare support. IEEE Access, 7: 185322–185337, 2019. doi: 10.1109/ACCESS.2019.2960615.

O. Ajayi, A. Bagula, and H. Maluleke. Africa 3: A continental network model to enable the african fourth industrial revolution. IEEE Access, 2020. doi:10.1109/ACCESS.2020.3034144.

O. Ajayi, A. Bagula, and H. Maluleke. The Fourth Industrial Revolution: A Technological Wave of Change. IntechOpen, 2022. doi: 10.5772/intechopen.106209.

D. Alaverdyen, F. Kuˇcera, and M. Horák. Implementation of the Smart City Concept in The EU: Importance of the Cluster initiatives and Best Practice Cases, pages 30–57. August 2018. doi: 10.2478/ijek-2018-0003.

V. Albino, U. Beradi, and R. Dangelico. Smart cities: Definitions, dimensions, performance, and initiatives. Journal of Urban Technology, 22(1):3–21, 2015.

L. E. Alfeld. Urban dynamics-the first fifty years. System Dynamics Review, 11(3):199–217, 1995.

M. Angelidou. The role of smart city characteristics in the plans of fifteen cities. Journal of Urban Technology, 24(2):19–46, August 2017. URL https://doi.org/10.1080/10630732.2017.1348880.

A. A. Aziza and T. D. Susanto. The smart village model for rural area (case study: Banyuwangi regency). In 3rd International Conference on Engineering & Technology for Sustainable Development (ICET4SD), volume 722, Oct. 2019. doi: 10.1088/1757-899X/722/1/012011.

T. B. Bayu. Smart leadership for smart cities. Smart Cities and Regional Development Journal, 4(1), 2020b. URL http://www.scrd.eu/index.php/scrd/article/view/71/63.

B. Benamrou, B. Mohamed, A. Bernouss, and O. Mustapha. Ranking models of smart cities. In 2016 10th International Conference on Complex, Intelligent, and Software Intensive Systems (CISIS), 2016. doi: 10.1109/CIST.2016.7805011.

J. Berthold and M. Hoglund Wetterwik. Examining the ecocity – from definition to implementation. In KTH Industrial Engineering and Management, Stockholm, 2013.

S. E. Bibri. Smart Sustainable Cities of the Future: The Untapped Potential of Big Data Analytics and Context–Aware Computing for Advancing Sustainability (The Urban Book Series). Springer, Switzerland, 1st edition, 2018a.

S. E. Bibri. The iot for smart sustainable cities of the future: An analytical framework for sensor-based big data applications for environmental sustainability. Sustainability of Cities and Society, 38:250–235, 2018b.

S. E. Bibri. Big data science and analytics for smart sustainable urbanism: unprecedented paradigmatic shifts and practical advancements. Springer Nature, 2019.

S. E. Bibri and J. Krogstie. On the social shaping dimensions of smart sustainable cities: Ict of the new wave of computing for urban sustainability. Sustainability of Cities and Society, 29:219–246, 2017.

CAICT and MIIT. Comparative Study of Smart Cities in Europe and China 2014. 2014.
CityPopulation. Congo democratic republic, 2020. URL https://www.citypopulation.de/en/drcongo/cities/.

B. Cohen. What exactly is a smart city?, 2012. URL https://www.fastcompany.com/1680538/what-exactly-is-a-smart-city.

B. Cohen. The smartest cities in the world 2015: Methodology. FastCompany, 11(20), 2014. URL https://www.fastcompany.com/3038818/the-smartest-cities-in-the-world-2015-methodology.

R. P. Dameri. Smart city implementation: creating economic and public value in innovative urban systems. Imprint, 2017.

UN DESA. World population prospects 2022: Summary of results, 2022. URL https://www.un.org/development/desa/pd/sites/www.un.org.development.desa.pd/files/wpp2022 summary of results.pdf.

J. Duran-Encalada and A. Paucar-Careres. System dynamics urban sustainability model for puerto aura in puebla, mexico. Systemic Practice and Action Research, 22(2):77–99, April 2009.

R. Dwevedi, K. Vinoy, and A. Kumar. Environment and big data: Role in smart cities of india. Resources, 7(64), 2018. doi: 10.3390/resources7040064.

Fajrillah, M. Zarina, and N. Wiria. Smart city vs smart village. Jurnal Mantik Penusa, 22(1):1–6, August 2018. doi: 10.31227/osf.io/r3j8z.

V. Fernandez-Anes, G. Velazquez, F. Perez-Prada, and A. Monzon. Smart city projects assessment matrix: Connecting challenges and actions in the mediterranean region. Journal of Urban Technology, 2018.

Matthias Finger. Smart city – hype and/or reality? IGLUS Q., 4(1):2–6, June 2018.

S. Gantori. How big is the smart opportunity in Asia?, page 15. 2019. URL https://www.ubs.com/content/dam/WealthManagementAmericas/cio-impact/Smart-Cities-7-March-2019.pdf.

R. Giffinger, C. Fertner, H. Kramar, R. Kalasek, N. Pichler-Milanovic, and E. Meijers. Smart cities. ranking of european medium-sized cities. Technical report, Vienna University of Technology, Vienna, 2007. URL http://www.smartcities.eu/download/smart-cities-final-report.pdf.

C. Giles. African "smart cities:" a high-tech solution to overpopulated megacities?, 2017. URL https://edition.cnn.com/2017/12/12/africa/africa-new-smart-cities/index.html.

A. Gyrard, A. Zimmermann, and A. Sheth. Building iot-based applications for smart cities: How can ontology catalogs help? IEEE Internet of Things Journal, 5(5), 2018. doi: 10.1109/JIOT.2018.2854278.

El Haj and D. Ait. The moroccan city and the need for a transformation in the era of the smart city: Analysis of the cases of the cities of tangier, casablanca and marrakesh. Geopolitics and Geostrategic Intelligence, 3(2): 66–84, November 2020.

Y. R. Jabareen. Sustainable urban forms: Their typologies, models, and concepts. Journal of Planning Education and Research, 2006. doi: 10.1177/0739456X05285119.

N. Komninos, C. Bratsas, C. Kakderi, and P. Tsarchopoulos. Smart city ontologies: Improving the effectiveness of smart city applications. Journal of Smart Cities, 1(1), 2019. doi: 10.18063/JSC.2015.01.001.

K. Laaboudi. Power of the people in smart cities, November 2016. URL http://www.e-madina.org/wp-content/uploads/2016/11/White-Paper-e-Madina-3.0-Value-Chain-of-Smart-cities.pdf.

S. Lamari and S. Oukarfi. Adoption and use of smartphones in casablanca: what impact on the casa smart-city project? Revue Géopolitique, Geostratégie et Intelligence Economique, 11:19–30, 2018.

D Lee. Africa is ready to leapfrog the competition through smart cities technology. Deloitte & Touche, 2014.

V. Manirakiza, L. Mugabe, A. Nsabimana, and M. Nzariyambah. City profile: Kigali, rwanda. Environment and Urbanization ASIA, 10(2):290–307, 2019a.

V. Manirakiza, L. Mugabe, A. Nsabimana, and M. Nzariyambaho. City profile: Kigali, rwanda. Environment and Urbanization ASIA, 10(2):290–307, 2019b.

C. Manville, G. Cochrane, J. Cave, J. Millard, J. Pederson, R. Thaarup, A. Liebe, M. Wissner, R. Massink, and B. Kotterink. Mapping smart cities in the eu, 2014. URL https://www.europarl.europa.eu/RegData/etudes/etudes/join/2014/507480/IPOL-ITRE ET(2014)507480 EN.pdf.

A. Martos, R. Pacheco-Torres, J. Ordonez, and E. Jadraque-Cago. Towards successful environmental performance of sustainable cities: Intervening sectors. a review. Renewable and Sustainable Energy Reviews, 57:479–495, May 2016. doi: 10.1016/j.rser.2015.12.095.

A. Maslon-Oracz and M. Mazurewicz. Smart regions and cities supporting cluster development and industrial competitiveness in the European Union. Africa's Smart City development influencing global competitiveness, pages 335–345. Warsaw, 2015.

M. Mekni and K. Huard. Smart cities in africa: examples of the cities of cairo and casablanca, 2020. URL https://www.bearingpoint.com/fr-fr/publications-evenements/blogs/energie/smart-cities-en-afrique-exemples-des-villes-du-caire-et-de-casablanca/.

Ministry of Land, Infrastructure and Transport (MLIT). Korean smart cities, 2019. URL http://smartcity.go.kr/wp-content/uploads/2019/08/Smart-city-broschureENGLISH.pdf.
T. Nam and T. A. Pardo. Conceptualizing smart city with dimensions of technology, people, and

institutions. In Proc. 12th Annual International Conference on Digital Government Research, pages 283–291, June 2011. doi: 10.1145/2037556.2037602.

G. Natarajan and L. Kumar. Implementation of iot based smart village for the rural development. International Journal of Mechanical Engineering and Technology, 8(8):1212–1222, 2017.

P. Neirotti, A. De Marco, A. Cagliano, G. Mangano, and F. Scorrano. Current trends in smart city initiatives: Some stylised facts. Cities, 38:25–36, 2014.

D. Nkurikiyimfura. Smart africa & smart cities initiative, 2016. URL http://media.firabcn.es/content/S078018/download/14NOV GF EGOV KN2.pdf.

N. Odendaal. Getting smart about smart cities in cape town. beyond the rhetoric. Smart urbanism: Utopian vision or false dawn, pages 71–87, 2016.

M. Pichler. Smart city vienna: System dynamics modelling as a tool for understanding feedbacks and supporting smart city strategies, 2017.

E. Pieterse and T. Zevi. Africa 2063: How cities will shape the future of a continent. 2018. URL https://www.ispionline.it/sites/default/files/pubblicazioni/ispi-dossier-pieterse-zevi-31.05.2019-0.pdf.

C. Rabari and M.l Storper. The digital skin of cities: urban theory and research in the age of the sensored and metered city, ubiquitous computing and big data. Cambridge Journal of Regions, Economy and Society, pages 27–42, 2015. doi: 10.1093/cjres/rsu021.

M. Reda Khomsi and B. Bedard. From smart city to smart destination. the case of three canadian cities. J. Tour. Res., 6(2):69–74, 2016.

R. Register. Visionary revolutionary, 2013. URL https://ecocitybuilders.org/richard-register/.

R. Rich, P. Westerberg, and J. Toner. Smart city rwanda masterplan. 2017. URL https://unhabitat.org/smart-city-rwanda-master-plan.

M. Roseland. Dimensions of eco-city. Cities, 14(4):197–202, 1997.

M. Roseland. The eco-city approach to sustainable development in urban areas? In How Green is the City? Sustainability Assessment & the Management of Urban Environments, pages 85–103. 2001.

Smartcity. The leading cities of africa – nairobi and cape town. 2018. URL: https://smartcity.press/nairobi-cape-town-smart-cities-africa/.

J. Tadili and H. Fasly. The digital transformation of the city or the smart city: Terminology and definitions. Revue Contrôle Comptabilité et Audit, 25(4): 277–296, 2019.

R. Teipelke. Urban planning realities of smart urban data applications. IGLUS-Q., 4(1):7–12, June 2018.

UBS. March 2019. URL https://www.ubs.com/content/dam/WealthManagementAmericas/cio-impact/Smart-Cities-7-March-2019.pdf.

UN. Sustainable development goals. URL https://www.un.org/sustainabledevelopment/sustainable-development-goals/.

UN. World urbanization prospects. 2014.

UN. The Sustainable Development Goals Report 2023 Special edition. Towards a Rescue Plan for People and Planet. 2023.

P. Varghese. Smart-cities for india: Why not open-source villages?in research into design for communities. In Proceedings of ICoRD 2017, pages 79–89, 2016a.

P. Varghese. Exploring other concepts of smart-cities within the urbanising indian context. Urban Research and Practice, 9(2):1858–1867, 2016b.

N. Viswanadham and S. Vedula. Design smart villages. The Centre for Global Logistics and Manufacturing Strategies Indian School of Business, pages 1–16, 2010.

T. Waas, A. Verbruggen, and T. Wright. University research for sustainable development: definition and characteristics explored. Journal of Cleaner Production, pages 629–636, 2010.

M. Weiser. The computer for the 21st century. Scientific American, 26(3): 94–104, 1991.

Wikipedia. Lubumbashi, 2020. URL https://fr.wikipedia.org/wiki/Lubumbashi.

F. Wolman. Urban Dynamics. The M.I.T. Press, Cambridge, 1969.

V. Zavratnik, D. Podjed, J. Trilar, N. Hlebec, A. Kos, and E. S. Duh. Sustainable and community-centred development of smart cities and villages. Sustainability, 12(10):3961, 2020.